\documentclass{article}
\usepackage{icrctc07}
\def \mal {Malarg\"{u}e}

\def \eev { \mathrm{\ EeV}}

\def \gcm {\mathrm{\ g\ cm^{-2}}}

\newcommand{\fluxunit}{km$^{-2}$~sr$^{-1}$~yr$^{-1}$}

\newcommand{\apertureunit}{km$^{2}$~sr~yr}

\newcommand{\xmax}{X$_{\rm max}$}

\title{Search for Ultra-High Energy Photons with the Pierre Auger Observatory}

\shorttitle{UHECR Photon Limit}

\authors {M. Healy$^{1}$ for the Pierre Auger Collaboration$^{2}$}
\shortauthors{M. Healy for the Pierre Auger Collaboration}
\afiliations{$^{1}$University of California, Los Angeles, Los Angeles,
CA 90095, USA\\$^{2}$Av. San Martin Norte 304 (5613) \mal, Prov. de Mendoza, Argentina}

\email{bes@ast.leeds.ac.uk}

\abstract{Data taken at the Pierre Auger Observatory are used to
search for air showers initiated by ultra-high energy (UHE)
photons. Results of searches are reported from hybrid observations
where events are measured with both fluorescence and array
detectors. Additionally, a more stringent test of the photon fluxes
predicted with energies above $10^{19}$~eV is made using a larger data
set measured using only the surface detectors of the observatory.}

\begin{document}
\maketitle
%Begin the section.

\section{Introduction}
It has been suggested that the excess of cosmic-rays with energies
above the predicted Greisen-Zatsepin-Kusmin (GZK) steepening, observed
by AGASA, could be due to cosmic-rays being produced as the
by-products of 'top-down' scenarios and not by more conventional
acceleration mechanisms, such as the diffusive shock process, in
astrophysical objects. If the former were the case, it would be
expected that a significant proportion of the spectrum of cosmic rays
at the highest energies would be UHE photons: predictions vary from
model to model but are in the range 10\% to 50\% above 10~$\eev$.  The
extensive air showers (EAS) generated by UHE photons reach shower
maximum \xmax, the atmospheric depth at which the number of charged
particles in the shower is greatest, at much greater depths than their
nuclear counterparts. This is due to the much lower multiplicity in
particle production in the electromagnetic-dominated photon showers
than in the hadronic interactions present for nuclear primaries. The
depth of maximum is further increased above 30~$\eev$ because of the
suppression of Bethe-Heitler pair production due to the LPM effect
\cite{LPMEffect_LP}\cite{LPMEffect_M}, which is not important for other cosmic-ray primaries.

\xmax~ can thus be used to discriminate between photon and nucleonic
UHE primaries. With the Pierre Auger Observatory this can be done
using hybrid events which are observed with both fluorescence and
array detectors, and in which \xmax~ is measured directly. This allows
the determination of a limit to the integral fraction of photons above
10~$\eev$. Parameters measured using only the surface array of
detectors, which reflect shower maximum, can also be compared to
predictions for photons from simulations to provide a stronger limit
on the flux of photons at several energies
\cite{SDParametersPhotonLimit}. Two such observables are used which
behave differently for nuclear primaries when compared to photons:
these are the risetime of the signal in the water-Cherenkov detectors
at 1000 m from the core and the radius of curvature of the shower
front.

\section{Analysis of Array Measurements}
Data were taken with the surface detector over the period 1 January
2004 to 31 December 2006 for showers with zenith angles of
30$^{\circ}$ to 60$^{\circ}$, corresponding to an aperture of
$3130$~\apertureunit{}. For each event, the risetime at
1000 m, $t_{1/2}(1000)$, and the radius of curvature $R_c$ was
found. Photon candidates were selected with a cut determined \emph{a
priori} with the characteristics that these parameters were expected to display
if they were photonic, derived using Monte Carlo simulations. Upper
limits were then calculated, both to the absolute flux and to the
fraction of photons from the number of photon candidates. The
motivation for using the chosen parameters is now discussed.

\begin{figure}[htbp]
  \vspace{-10pt}
  \centering
  %\subfigure[] {
    %\centering
    \includegraphics*[width=0.45\textwidth]{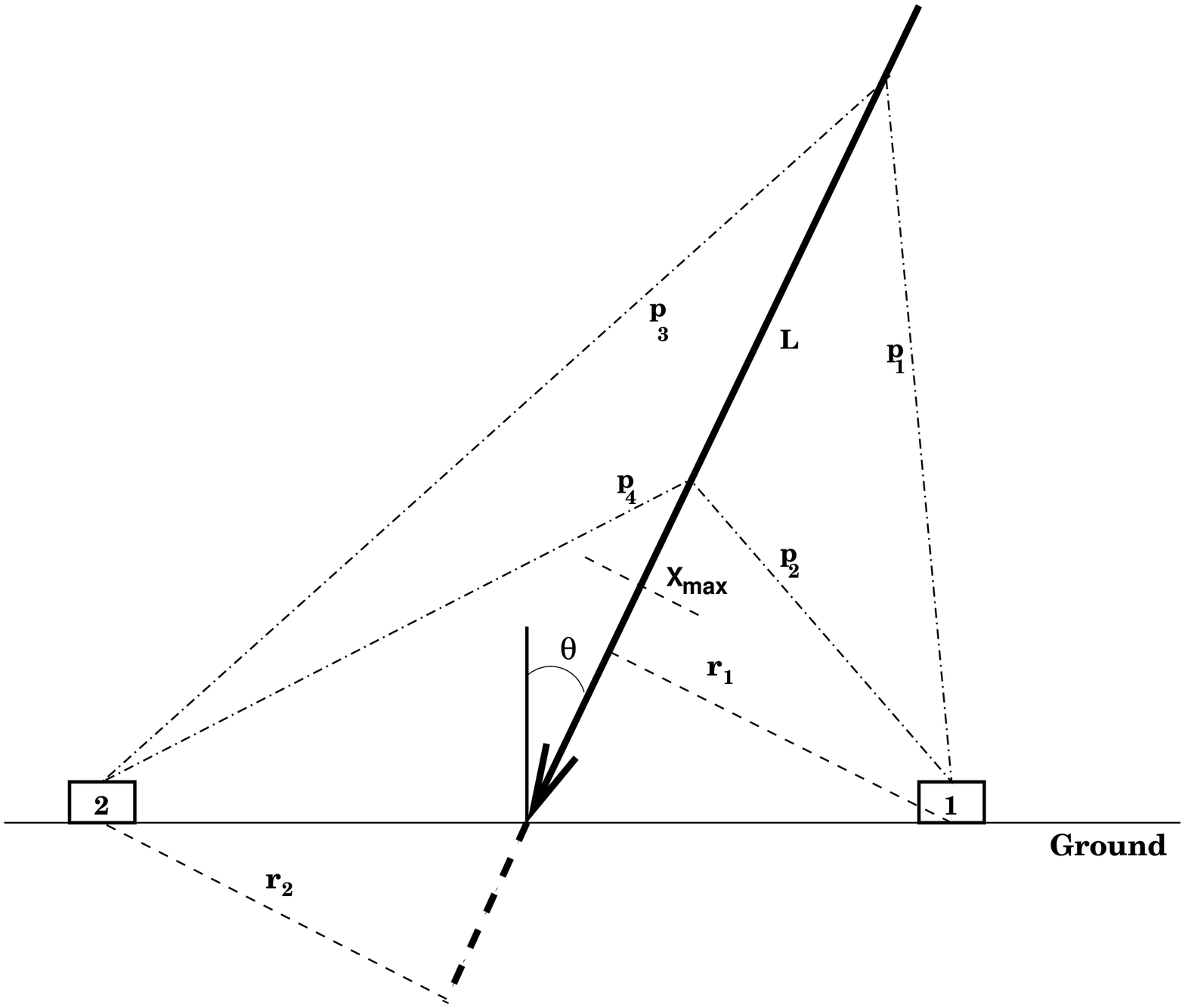}
        
  %} %\subfigure[] { %\centering
    \includegraphics*[width=0.45\textwidth]{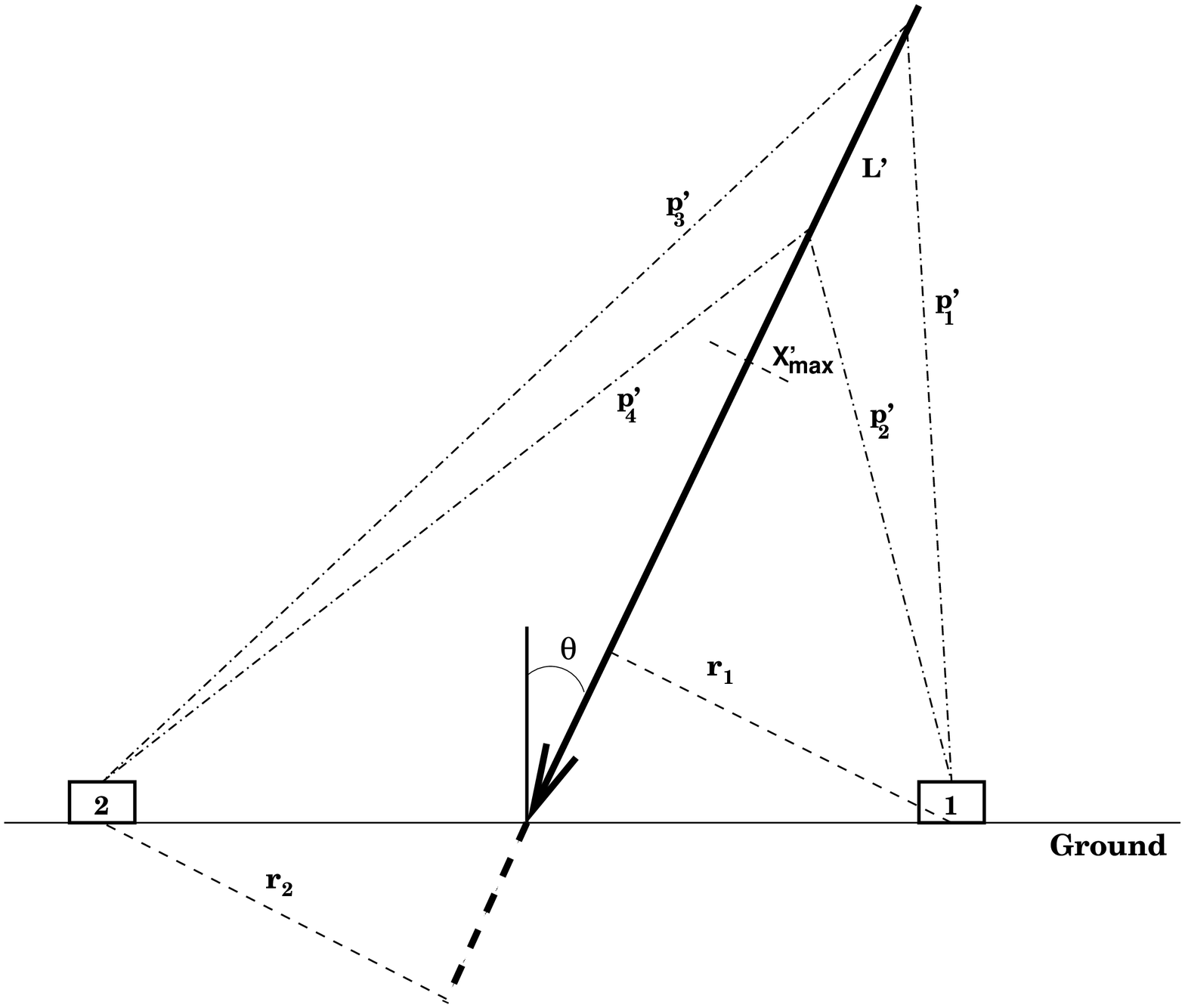}
        
  %}
    \caption{Example of the difference in path lengths for particles
    arriving at detectors of an EAS with different \xmax. The top
    drawing shows a deeply penetrating shower while on the bottom is a
    shallow shower (\xmax $>$ X'$_{\rm max}$). The difference in paths
    is larger for the deeper shower resulting in greater time
    spreads. Although r$_{\rm 1}$ $=$ r$_{\rm 2}$, station 2 samples
    the shower later in the development than station 1 introducing an
    asymmetry in the measurement of $t_{1/2}$.  }
  \label{RisetimeDemonstration}
\end{figure}

\subsection{Risetime}

The arrival time distribution of particles in showers is expected to
become more dispersed with increasing \xmax~due to the geometry
associated with particles in the shower: the difference in path
lengths between particles produced early and late in the shower is
greater for EAS which penetrate deeper into the atmosphere as
demonstrated in figure \ref{RisetimeDemonstration}. Photon showers
dissipate energy over larger atmospheric track lengths than their
nuclear counterparts, reaching \xmax~ deeper in the atmosphere and
increasing the spread of the time signals relative to hadronic
primaries.

The signal time spreads are evaluated by defining the risetime,
$t_{1/2}$, as the time it takes for the signal to rise from 10\% to
50\% of the total signal deposited in the array detectors. Each
measurement of $t_{1/2}$ is corrected for asymmetry effects: stations
which lie at similar core distances in the same event will sample the
shower at different stages of development, due to the different
azimuthal directions of the stations relative to the shower
plane. This is demonstrated in figure \ref{RisetimeDemonstration},
where station 2 samples the shower later in the development than
station 1, resulting in a shorter signal despite the core distances
being identical.

On an event-by-event basis, the estimated risetime at 1000 m from the
core, $t_{1/2}(1000)$ is found by fitting the risetimes from
individual stations as a function of core distance. Due to the much
larger \xmax~ expected from photons than nuclei, one would expect
larger $t_{1/2}(1000)$ for photons than from proton or iron primaries.

\subsection{Radius of Curvature}
The early stages of a shower include the production of high energy
muons which are relatively unscattered as they propagate to the
ground. Accordingly they form a shower front which expands in an
approximately spherical manner, where the radius of curvature of this
sphere, $R_{c}$, is related to the distance of the observer from the
production region of the muons, which in turn relates to \xmax. One
would thus expect larger values of \xmax~ to be reflected by smaller
values of $R_{c}$.

\section{Results and Discussion}
\subsection{Hybrid Results}
\begin{figure}[htbp]
\vspace{-10pt}
\begin{center}
\includegraphics[width=0.45\textwidth]{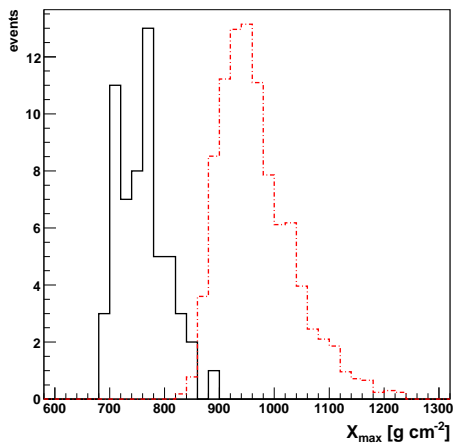}
\end{center}
\caption[font=small,labelfont=bf, aboveship=2pt]{Black: The
distribution of \xmax from 58 hybrid events with energies above
10~$\eev$ that meet selected cuts\cite{OriginalHybridPhotonLimit}
. The dashed line shows a distribution for 10~$\eev$ photons arriving
over a range of zenith angles.  All events reach shower maximum at
depths which are too shallow to be considered as a result of photon
primaries.}
\label{HybridXmaxDistribution}
\end{figure}

In \cite{OriginalHybridPhotonLimit} we reported a limit to the
fraction of photons in the integral cosmic-ray flux of 16\% (95\%
c.l.) above 10~$\eev$ based on 29 high-quality hybrid events recorded
in the period Jan. 2004 - Feb. 2006. We have updated the analysis with
data collected until March 2007, keeping the analysis cuts as in the
original paper. In total, 58 events are now available. The measured
\xmax~distribution is shown in figure \ref{HybridXmaxDistribution}
along with the calculated distribution from 10~$\eev$ photons made on
an event-by-event basis. Even the largest observed value of \xmax,
$\sim$~900~$\gcm$, is well below the average value expected for
photons (about $\sim$~1000~$\gcm$, see e.g. Table 1 in
\cite{OriginalHybridPhotonLimit}). With the updated sample, the upper
limit becomes 13\% (95\% c.l.) above 10~$\eev$.

\begin{figure}[htbp]
\vspace{-0pt}
\begin{center}
\includegraphics[width=0.45\textwidth, height=0.4\textwidth]{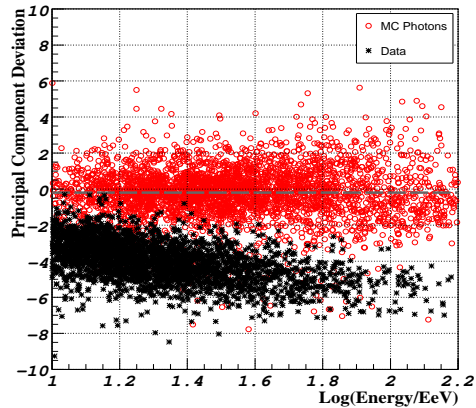}
\end{center}

\caption[font=small,labelfont=bf, aboveship=2pt]{The principle
component between risetime and radius of curvature as a function of
energy for data (black) and simulated photons (red). Data lying above
the dashed line, which is the mean of the distribution for photons,
are identified as photon candidates. No events meet this requirement.}
\label{UCLAPrincipleComponentPlot}
\end{figure}

\subsection{Surface Detector Results}

Photon candidates were searched for by combining the risetime and
radius of curvature for each event into one observable
using a principle component analysis, where the angle of rotation is
found using 5\% of the data. The remaining 95\% of showers were then
identified as nuclei or photon candidates using an \emph{a priori}
cut, where showers are excluded as photon-like if the principle
component measurement is less than the mean of that predicted by a
spectrum of photonic showers generated by Monte Carlo
simulations. Events which were deemed as photonic were assigned
energies using a reconstruction designed from photon simulations,
reflecting the fact that these showers have low muon content and
are more strongly attenuated than hadronic showers
\cite{Billior}. The spectrum of simulated photons incorporates the
efficiency of photon detection and reconstructions, as
well as including the possibility that photons may interact with the
geomagnetic field before arriving in the atmosphere.

\begin{figure}[htbp]
  \vspace{-120pt}
  \subfigure[Upper limits to photon fractions] {
        \centering
        \includegraphics[width=0.45\textwidth]{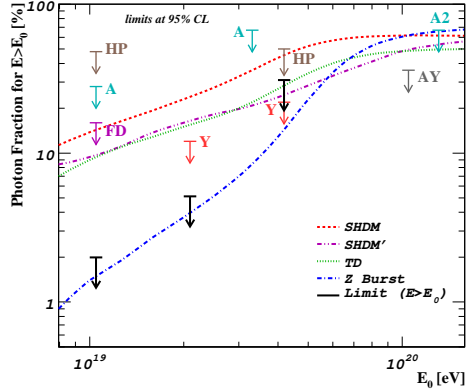}
        %\label{fig:subfig1}
    }
    \subfigure[Upper limits to photon flux] {
      \centering
      \includegraphics[width=0.45\textwidth]{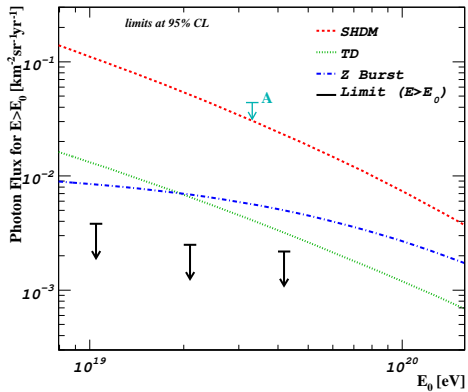}
        %\label{fig:subfig2}
    }
    \caption[font=small,labelfont=bf, aboveship=2pt]{(a) The upper limits to the fraction of photons including limits from previous experiments and predictions from top-down models based on the AGASA spectrum. The limits from this work are shown in black. The limit derived from the hybrid data is also shown and labelled FD. Labels A, HP and Y refer to the limits set by AGASA, Haverah Park and Yakutsk, for references see \cite{review}. (b) The upper limits to the flux of photons along with predictions of top-down models.}
  \label{ModelsVsLimits}
\end{figure}

The results are summarised in figure \ref{UCLAPrincipleComponentPlot}
where the principle component is plotted for data (black) and the
simulated photons (red) as a function of energy. There are 0 events
above the dashed line, where the cut is defined, and as such there are
no candidate photons for events above 10 $\eev$. This constrains the
maximum flux of photons to $3.8 \times 10^{-3}$,~$2.5 \times
10^{-3}$~and $2.2 \times 10^{-3}$~\fluxunit{} above 10, 20 and 40
$\eev$ respectively. The upper limit on the photon fraction, based on
the measured spectrum of events in \cite{SommersSpectrum}, is
calculated as 2.0\%, 5.1\% and 31\% at 10, 20 and 40 $\eev$
respectively. These limits are shown against the predictions for
photons from top-down models based on the AGASA spectrum (see
\cite{review}) in figure \ref{ModelsVsLimits}. Also shown are upper
limits to the flux and fractions of photons from previous
experiments. This work constrains the photon fractions and fluxes to
more stringent limits than previously measured and disfavours the
proposed top-down models as the sources of UHECRs and of the AGASA
events.

On completion of the surface array, the Pierre Auger Observatory will reach
sensitivities of $4 \times 10^{-4}$~\fluxunit{} for the integrated
flux and 0.7\% for the fraction of photons above 20 $\eev$ (95\% c.l.)
after 5 years of operation.

%\bibliographystyle{plain}

%\begin{thebibliography}{99}

%\bibitem{OriginalHybridPhotonLimit}
%Pierre Auger Coll., Astropart. Phys. 27 (2007) 155-168

%\bibitem{SDParametersPhotonLimit}
%X.~Bertou, P.~Billoir, S.~Dagoret-Campagne,
%Astropart.~Phys. {\bf 14}, 121 (2000).

%\bibitem{LPMEffect} 
%L.D.~Landau, I.Ya.~Pomeranchuk,
%Dokl. Akad. Nauk SSSR {\bf 92}, 535 \& 735 (1953);
%A.B.~Migdal, Phys. Rev. {\bf 103}, 1811 (1956).

%\bibitem{review}
%M. Risse and P. Homola, Mod. Phys. Lett. A 22, 749 (2007).

%\bibitem{TopDownModels} 
%G.~Gelmini, O.E.~Kalashev, D.V.~Semikoz, [arXiv:astro-ph/0506128].

%\bibitem{SHDMModels}
%J.~Ellis, V.~Mayes, D.V.~Nanopoulos, Phys.~Rev.~D {\bf 74}, 115003 (2006).

%\bibitem{Billior} 
%P.~Billoir, C.~Roucelle, J.C.~Hamilton, [arXiv:astro-ph/071583]

%\bibitem{SommersSpectrum}
%P.~Sommers for the Pierre Auger Coll., Proc. 29th ICRC, Pune (2005), ICRC-05-124, [arXiv:astro-ph/0507150]

%\end{thebibliography}

%This is the reference to .bib file (Whitout .bib!)
\bibliography{icrc0602}
%This in the bibtex style, is ok.
\bibliographystyle{plain}
\end{document}